\documentclass[11pt]{article}
\usepackage{epsfig,amssymb,bm} 
\newcommand{\cD}{{\cal D}} 
\newcommand{\cN}{{\cal N}} 
\newcommand{\cO}{{\cal O}}

\def\bbbone{{\rm 1\hspace{-1.1mm}I}}	

\newcommand{\rd} {\mathrm d}
\newcommand{\re} {\mathrm e}
\newcommand{\ri} {\mathrm i}
\newcommand{\vc} {{\bm c}}
\newcommand{\vk} {{\bm k}}

\newcommand{\nn}{\nonumber} 
\newcommand{\be}{\begin{equation}} 
\newcommand{\ee}{\end{equation}}  
\newcommand{\bea}{\begin{eqnarray}}
\newcommand{\eea}{\end{eqnarray}}

\begin{document} 
\title{\Large\bf Spectra of Modular and Small-World Matrices} 

\author{Reimer K\"uhn$^1$ and Jort van Mourik$^2$\\
$^1$Mathematics Department, King's College London, Strand, London WC2R 2LS,UK\\
$^2$NCRG, Aston University, Aston Triangle,  Birmingham B4 7ET, UK}

\maketitle

\begin{abstract}
We compute spectra of symmetric random matrices describing graphs with {\em general\/} 
modular structure and  arbitrary inter- and intra-module degree distributions, subject 
only to the constraint of finite mean connectivities. 
We also evaluate spectra of a certain class of small-world matrices generated from
random graphs by introducing short-cuts via additional random connectivity components. 
Both adjacency matrices and the associated graph Laplacians are investigated.
For the Laplacians, we find Lifshitz type singular behaviour of the spectral density in 
a localised region of small $|\lambda|$ values.
In the case of modular networks, we can identify contributions local densities of state 
from individual modules.
For small-world networks, we find that the introduction of short cuts can lead to the
creation of satellite bands outside the central band of extended 
states, exhibiting only localised states in the band-gaps. Results for the ensemble in 
the thermodynamic limit are in excellent agreement with those obtained via a cavity 
approach for large finite single instances, and with direct diagonalisation results.
\end{abstract}
\section{Introduction}
\label{sec:intro}
The past decade has seen a considerable activity in the study of random graphs
(see, e.g. \cite{New+01}, or \cite{AlbBarab02, New03, DorogMend03} for recent 
reviews), as well as concurrent intensive studies in spectral properties of sparse 
random matrices \cite{AlbBarab02, Dorog+03, Cvetk+95, Bollobas01, Farkas+01}, the 
latter providing one of the key tools to study properties of the former. Moments of 
the spectral density of an adjacency matrix describing a graph, for instance give 
complete information about the number of walks returning to the originating vertex 
after a given number of steps, thus containing information about local topological 
properties of such graphs. Spectral properties, specifically properties of eigenvectors 
corresponding to the largest eigenvalue of the modularity matrices of a graph and 
of its subgraphs -- matrices closely related to the corresponding adjacency matrices
-- can be used for efficient modularity and community detection in networks 
\cite{New06}, and so on. Much of this activity has been motivated by the fact that a 
large number of systems, natural and artificial, can be described using network 
descriptions of underlying interaction patterns, and the language and tools of graph 
theory and random matrix theory for their quantitative analysis.

Though the study of spectral properties of sparse symmetric matrices was initiated 
by Bray and Rodgers already in the late 80s \cite{RodgBray88, BrayRodg88}, fairly 
complete analytic and numerical control over the problem has emerged only recently 
\cite{Ku08, Rog+08}, effectively using generalisations of earlier ideas developed by  
Abou-Chacra et al. \cite{AbouCh+73} for Bethe lattices. Analytical results for spectral 
properties of sparse matrices had typically been based either on the single defect or 
effective medium  approximations (SDA, EMA) \cite{BirMon99,SemerjCugl02, Dorog+03, 
NagRog08}, or were restricted to the limit of large average connectivity \cite{Rodg+05, 
KimKahng07}. Alternatively, spectra for systems with heterogeneity induced by 
scale-free or small-world connectivity \cite{Farkas+01, Goh+01}, or as a result 
of an explicitly modular structure \cite{MitTad08} were obtained through numerical 
diagonalisation. Analytical results for spectra of modular systems \cite{ErgKu09} 
and for systems with topological constraints beyond degree distributions 
\cite{Rog+10} are still very recent.

The purpose of the present paper is to expand the scope of \cite{ErgKu09} in two ways, 
(i) by providing spectra of random matrices describing graphs with {\em general\/} 
modular structure and arbitrary inter- and intra-module degree distributions, 
subject only to the constraint of finite mean connectivities, and (ii) by computing 
spectra for a class of small-world systems, constructed as regular random graphs
with an additional connectivity component providing long-range interactions and thus 
short-cuts. The connection between these two seemingly different problems is mainly 
provided by the close similarity of the methods used to study these systems.

Our study is motivated by the fact that modularity of systems, and thus networks of 
interactions is a natural property of large structured systems; think of 
compartmentalisation in multi-cellular organisms, sub-structures and organelles 
{\em inside\/} cells and the induced structures e.g. in protein-protein interaction 
networks, or think of large corporates with several subdivisions, to name but a few 
examples.

In Sect. \ref{sec:setup} we introduce the type of multi-modular system and the associated
random matrices we are going to study. A replica analysis of the problem is described in
Sect. \ref{sec:repl}, with (replica-symmetric) self-consistency equations derived in
Sect. \ref{sec:rs}.  Sect. \ref{sec:smw} introduces a class of small-world networks 
generated from (regular) random graphs by introducing short-cuts via a second,
long-range  connectivity component, and briefly describes the rather minimal 
modifications in the theoretical description needed to analyse those systems as well. 
In Sect. \ref{sec:results} we present a selection of results. Our main conclusions are
outlined in Sect. \ref{sec:concl}.

\section{Modular Systems}
\label{sec:mod}
\subsection{Multi-Modular Systems and Random Matrices Associated with Them}
\label{sec:setup}
We consider a system of size $N$ which consists of $m$ modules $\cN_\mu$, 
$\mu=1,\dots,m$.
We use $N_\mu$ to denote the size of the module $\cN_\mu$, and assume that each module 
occupies a finite fraction of the entire system, $N_\mu= f_\mu N$, with $f_\mu > 0$ for
all $\mu$, and
\be
\sum_{\mu=1}^m f_\mu = 1\ .
\ee 
Details of the modular structure are encoded in the $N\times N$ connectivity matrix
$\vc =(c_{ij})$, whose matrix elements $c_{ij}$ describe whether a link between nodes $i$ 
and $j$ exists ($c_{ij}=1$) or not ($c_{ij}=0$). To each site site $i$ of the system, 
we assign a connectivity vector $\vk_i=(k_i^\nu)$, whose components 
\be
k_i^\nu=\sum_{j\in\cN_\nu} c_{ij}
\ee
give the number of connections between site $i$ and (other) sites in module $\nu$. 
The $\vk_i$ are taken to be fixed according to some given distribution which we assume 
to depend only on the module to which $i$ belongs, and which has finite 
means 
\be
\langle k_i^\nu\rangle_\mu = c^{\mu\nu}\ ,\qquad i\in \cN_\mu
\ee
for the components, but is otherwise arbitrary. We use $\langle \dots\rangle_\mu$ to 
denote an average over the distribution of coordinations for vertices in module $\cN_\mu$. 
Consistency required by symmetry entails $c^{\mu\nu}/N_\nu= c^{\nu\mu}/N_\mu$, or alternatively  
$f_\mu c^{\mu\nu} =  f_\nu c^{\nu\mu}$.

Starting from the modular structure defined by the connectivity matrix $\vc$, we consider
two types of random matrix inheriting the modular structure. The first is defined
by giving random weights to the links, thereby defining random matrices $M$ of the form
\be
M_{ij} = c_{ij}  K_{ij}\ ,
\label{M1}
\ee
where we assume that the statistics of the $K_{ij}$ respects the modular structure defined
by $\vc$ in that it only depends on the modules to which $i$ and $j$ belong. The second is 
related to the first by introducing zero row-sum constraints, resulting in matrices of the
form
\be
L_{ij} = c_{ij}K_{ij} - \delta_{ij} \sum_k c_{ik}K_{ik}\ .
\label{ML}
\ee
In the special case $K_{ij}={\rm const.}$, one recovers the connectivity matrices 
themselves, and the discrete graph Laplacians respectively.

We note in passing that it is possible to include extensive intra-module and inter-module 
connections in addition to the finite connectivity structure described above as in 
\cite{ErgKu09}, but we have decided not to do so here.

The spectral density of a given matrix $M$ can be computed from its resolvent via
\bea
\rho_M(\lambda) &=& \lim_{\varepsilon\searrow 0} ~ \frac{1}{\pi}{\rm Im}\,{\rm Tr} \left 
[\lambda_\varepsilon\bbbone -M \right]^{-1}\nn\\
&=&\lim_{\varepsilon\searrow 0} ~ \frac{-2}{N\pi}{\rm Im} 
\frac{\partial}{\partial \lambda} \ln {\rm det}\left[\lambda_\varepsilon\bbbone
-M \right]^{-1/2}\ ,
\label{defdos}
\eea
in which $\lambda_\varepsilon\equiv\lambda- i\varepsilon$, and the inverse square 
root of the determinant is obtained as a Gaussian integral. We are interested in the
average spectral density obtained from (\ref{defdos}) by taking an average over the 
ensemble of matrices considered, thus in
\be
\rho(\lambda)= \lim_{\varepsilon\searrow 0} ~ \frac{-2}{N\pi}{\rm Im} 
\frac{\partial}{\partial \lambda} \left \langle \ln \left[
\int \prod_i \frac{\rd u_i}{\sqrt{2\pi/\ri}} \exp\left\{- \frac{\ri}{2} \sum_{ij} u_i
\left[\lambda_\varepsilon \delta_{ij} -M_{ij}\right] u_j\right\}\right] \right\rangle\ ,
\label{avdos}
\ee
where angled brackets on the r.h.s denote an average over connectivities $\{c_{ij}\}$ and
weights $\{K_{ij}\}$ of the non-vanishing matrix elements. For the ensembles considered 
here the spectral density is expected to be self-averaging, i.e. that (\ref{defdos}) and
(\ref{avdos}) agree in the thermodynamic limit $N\to \infty$.

The distribution of connectivities is taken to be maximally random compatible with the 
distribution of coordinations. Vertices $i\in\cN_\mu$ and $j\in\cN_\nu$ are connected 
with a probability proportional to $k_i^\mu k_j^\nu$. This can be expressed in terms of a 
fundamental distribution $p^{\mu\nu}_0$ of connectivities (between sites $i\in\cN_\mu$ 
and $j\in\cN_\nu$)
\be
p^{\mu\nu}_0(c_{ij})=\left(1-\frac{c^{\mu\nu}}{N_\nu}\right) \delta_{c_{ij},0} + 
\frac{c^{\mu\nu}}{N_\nu} \delta_{c_{ij},1}\ .
\ee
as
\bea
P(\bm c|\{\vk_i\}) &=& \frac{1}{\cN}\ 
\prod_\mu 
\left\{
\Big\{\prod_{i<j \in \cN_\mu} p^{\mu\mu}_0(c_{ij}) \delta_{c_{ij},c_{ji}} \Big\}
\times \prod_{\nu(>\mu)}
\Big\{\prod_{i\in\cN_\mu}\prod_{j\in\cN_\nu} p^{\mu\nu}_0(c_{ij}) \delta_{c_{ij},c_{ji}} \Big\}
\right\}\nonumber\\
 & &\times \prod_{\mu} \prod_{i\in\cN_\mu} \left\{ \prod_{\nu} \delta_{\Sigma_{j\in\cN_\nu}
c_{ij}, k_i^\nu}
\right\}
\eea
where $\cN$ is a normalisation constant, and the Kronecker-deltas enforce
the coordination distributions.

The average of the logarithm in (\ref{avdos}) is evaluated using replica.
\be
\overline{\rho(\lambda)}= \lim_{\varepsilon\searrow 0} ~ \frac{-2}{N\pi}{\rm Im} 
\frac{\partial}{\partial \lambda}\lim_{n\to 0}\frac{1}{n} \ln \langle Z_N^n \rangle\ ,
\label{AvDOSn}
\ee
with
\be
Z_N^n= \int \prod_{ia} \frac{\rd u_{ia}}{\sqrt{2\pi/\ri}} \exp\left\{- \frac{\ri}{2} \sum_a
\sum_{ij} u_{ia} \left[\lambda_\varepsilon \delta_{ij} -M_{ij}\right] u_{ja}\right\}
\ee
Here $a=1,\dots,n$ enumerates the replica. We initially describe the process for 
matrices of type (\ref{M1}), and briefly mention the modifications to treat matrices  of
type (\ref{ML}) with zero row-sum constraints later.

\subsection{Disorder Average}
\label{sec:repl}
To evaluate the average, one uses integral representations of the Kronecker-deltas
\be
\delta_{\Sigma_{j\in\cN_\nu}c_{ij}, k_i^\nu} = \int \frac{\rd \varphi_i^\nu}{2\pi}\ 
\re^{\ri \varphi_i^\nu \left(\Sigma_{j\in\cN_\nu}c_{ij} - k_i^\nu\right)}
\ee

The average of the replicated partition function for matrices of type (\ref{M1}) becomes
\bea
\langle Z_N^n \rangle &=&\frac{1}{\cN}\  
\int \prod_{i\nu} \frac{\rd \varphi_i^\nu}{2\pi}\prod_{ia} \frac{\rd u_{ia}}{\sqrt{2\pi/\ri}} 
\exp\left\{-\frac{\ri}{2}  \lambda_\varepsilon \sum_{ia} u_{ia}^2 - \ri\sum_{\mu\nu} \sum_{i\in N_\mu}  
\varphi_i^\nu k_i^\nu \right .\nonumber\\
&& \left .
 + \sum_\mu \frac{c^{\mu\mu}}{2N_\mu}\sum_{i\ne j\in \cN_\mu}\left(~\Big\langle\exp\Big\{\ri K
\sum_{a} u_{ia}u_{ja}\Big\}\Big\rangle_{\mu\mu} \re^{\ri \varphi_i^\mu + \ri \varphi_j^\mu}   -1 \right)
\right .\nonumber\\
&& \left .
 + \sum_{\mu<\nu} \frac{c^{\mu\nu}}{N_\nu}\sum_{i\in \cN_\mu}\sum_{j\in \cN_\nu} \left( 
~\Big\langle\exp\Big\{\ri K \sum_{a} u_{ia}u_{ja}\Big\}\Big\rangle_{\mu\nu} 
\re^{\ri \varphi_i^\nu + \ri \varphi_j^\mu}   -1 \right) 
\right\}
\label{eq:avznd}
\eea
where $\langle ~\dots~\rangle_{\mu\nu}$ represents an average over the $K_{ij}$ 
distribution, connecting vertices $i\in\cN_\mu$ and $j\in\cN_\nu$, which is as yet left open.

Decoupling of sites is achieved by introduction of the replicated `densities'
\be
\rho^{(\mu\nu)}(\bm u,\varphi) =\frac{1}{N_\mu}
\sum_{i\in\cN_\mu} \prod_a \delta\Big( u_a - u_{ia}\Big) \delta\Big(\varphi-\varphi_i^\nu\Big)
\ee
and their $\varphi$ integrated versions
\be
\rho^{(\mu\nu)}(\bm u) = \int \rd\varphi\;\rho^{(\mu\nu)}(\bm u,\varphi) \re^{\ri \varphi}\ .
\label{udens}
\ee
It turns out that only the latter, and their conjugate densities $\hat\rho^{(\mu\nu)}$ are
needed, and Eq. (\ref{eq:avznd}) can be expressed as a functional integral
\be
\langle Z_N^n \rangle = \frac{1}{\cN}\ \int \prod_{\mu\nu} \{\cD\rho^{(\mu\nu)}\cD \hat\rho^{(\mu\nu)}\} 
\exp\left\{N\left[G_1 + G_2 + G_3\right]\right\}\ ,
\label{eq:avznpathint}
\ee
with
\bea
G_1 &=& \frac{1}{2} \sum_{\mu\nu} f_\mu c^{\mu\nu} \left( \int \rd \rho^{(\mu\nu)}(\bm u) 
\rd \rho^{(\nu\mu)}(\bm v) \Big\langle\exp\Big\{i K \sum_{a} u_{a} v_{a}\Big\}\Big\rangle_{\mu\nu}  -1\right)
\label{G1}\\
G_2 &=&  -\ri \sum_{\mu\nu} f_\mu \int \rd \bm u\,\hat\rho^{(\mu\nu)}(\bm u)\rho^{(\mu\nu)}(\bm u)
\label{G2}\\
G_3 &=& \sum_\mu  f_\mu\left\langle\ln \int \prod_a \frac{\rd u_a}{\sqrt{2\pi/\ri}} 
\prod_\nu \frac{\Big(\ri\hat\rho^{(\mu\nu)}(\bm u)\Big)^{k^\nu}}{k^\nu !}\,
\exp\left\{ - \ri\frac{\lambda_\varepsilon}{2} \sum_{a}  u^2_a \right\}
\right\rangle_{\mu}
\label{G3}
\eea
Here we have exploited the symmetry relation $f_\mu c^{\mu\nu} =  f_\nu c^{\nu\mu}$, 
introduced the short-hand notations $\rd \rho^{(\mu\nu)}(\bm u) \equiv
\rd \bm u \,\rho^{(\mu\nu)}(\bm u)$ for integrals over densities where appropriate, and
$\langle \dots\rangle_{\mu}$ in (\ref{G3}) for the average over the distribution of 
coordinations of sites in module $\cN_\mu$.

\subsection{Replica Symmetry and Self-Consistency Equations}
\label{sec:rs}
The functional integral (\ref{eq:avznpathint}) is evaluated by the saddle point method.
As in the extensively cross-connected case, the saddle point for this problem is expected
to be both replica-symmetric, and rotationally symmetric in the replica space. 
In the present context this translates to an ansatz of the form 
\bea
\rho^{(\mu\nu)}(\bm u) &=& \rho_0^{(\mu\nu)} \int \rd \pi^{(\mu\nu)}(\omega) \prod_a 
\frac{\exp\big[-\frac{\omega}{2} u_a^2\big]}{Z(\omega)}\ ,
 \nonumber\\
\hat\rho^{(\mu\nu)}(\bm u) &=& \hat \rho_0^{(\mu\nu)} \int \rd \hat\pi^{(\mu\nu)}
(\hat{\omega}) 
\prod_a \frac{\exp\big[-\frac{\hat\omega}{2}u_a^2 \big]}{Z(\hat{\omega})}\ ,
\label{RS}
\eea
with normalisation constants
\be
Z(\omega) =  \int \rd u \exp\left[-\frac{\omega}{2} u^2\right]
= \sqrt{2 \pi/\omega}\ ,
\label{defZ}
\ee
i.e. an uncountably infinite superposition of complex Gaussians (with $\mathrm{Re}
[\omega]\ge 0$ and $\mathrm{Re}[\hat\omega] \ge 0$) for the replicated densities and 
their conjugates \cite{Ku08, ErgKu09}. The $\rho_0^{(\mu\nu)}$,  $\hat\rho_0^{(\mu\nu)}$
in the expressions for $\rho^{(\mu\nu)}$ and $\hat \rho^{(\mu\nu)}$ in (\ref{RS}) are
determined such that the densities $\pi^{(\mu\nu)}$ and $\hat\pi^{(\mu\nu)}$ are 
normalised.

This ansatz translates path-integrals over the replicated densities $\rho^{(\mu\nu)}$ and 
$\hat\rho^{(\mu\nu)}$ into path-integrals over the densities $\pi^{(\mu\nu)}$ and $\hat\pi^{(\mu\nu)}$, 
and integrals over the normalisation factors $\rho_0^{(\mu\nu)}$  and $\hat \rho_0^{(\mu\nu)}$, giving
\be
\langle Z_N^n \rangle= \frac{1}{\cN}\ \int \prod_{\mu\nu} \{\cD\pi^{(\mu\nu)}\cD \hat\pi^{(\mu\nu)} 
\rd \hat \rho_0^{(\mu\nu)} \rd \rho_0^{(\mu\nu)}\} \exp\left\{N \left[G_1 + G_2 + G_3\right]\right\}\ ,
\label{ZnPint}
\ee
with
\bea
G_1 &\simeq& \frac{1}{2}\sum_{\mu\nu} f_\mu c^{\mu\nu}\Bigg[ \Big(\rho_0^{(\mu\nu)}\rho_0^{(\nu\mu)} 
- 1 \Big)\ ,\nn\\
& & + n\ \rho_0^{(\mu\nu)}\rho_0^{(\nu\mu)} \int \rd \pi^{(\mu\nu)}(\omega) \rd \pi^{(\nu\mu)}(\omega')  
\Bigg\langle \ln  \Bigg[\frac{Z_2( \omega,\omega',K)}{Z(\omega)Z(\omega')}\Bigg] \Bigg\rangle_{\mu\nu}\Bigg]
\label{G1n}\\
G_2 &\simeq& - \sum_{\mu\nu}f_\mu\, \ri \hat \rho_0^{(\mu\nu)} \rho_0^{(\mu\nu)} \Bigg[1 +
 n \int \rd \hat\pi^{(\mu\nu)}(\hat \omega) 
\,\rd \pi^{(\mu\nu)}(\omega) \ln \Bigg[\frac{Z(\hat{\omega}+\omega)}
{Z(\hat{\omega})Z(\omega)}\Bigg]\Bigg]\ ,
\label{G2n}\\
G_3 &\simeq& \sum_{\mu}  f_\mu\Bigg[\sum_\nu \Big( \Big\langle k^\nu \ln \ri \hat \rho_0^{(\mu\nu)} 
\Big\rangle_\mu - \Big\langle \ln k^\nu !\Big\rangle_\mu \Big )+ n \Bigg \langle \int \{\rd \hat\pi^{(\mu)}\}_{\bm k} 
\ln \Bigg[\frac{Z_\lambda(\Sigma_{\bm k}\,\hat\omega)}{\{Z\}_{\bm k}}\Bigg]
\Bigg\rangle_{\mu}\Bigg]\ .
\label{G3n}
\eea
Here, we have introduced short-hand notations for products of integration measures:  
$\{\rd\hat\pi^{(\mu)}\}_{\bm k}\equiv\prod_\nu\prod_{\ell_\nu=1}^{k^\nu}\rd
\hat\pi^{(\mu\nu)}(\hat\omega_{\ell_\nu})$, 
for products of partition functions: 
$\{Z\}_{\bm k} \equiv \prod_\nu \prod_{\ell_\nu=1}^{k^\nu} Z(\hat \omega_{\ell_\nu})$, 
and for $\hat \omega$-sums:
$\Sigma_{\bm k}\,\hat\omega\equiv \sum_\nu\sum_{\ell_\nu=1}^{k^\nu}
{\hat\omega}_{\ell_\nu}$ . 
Furthermore, we have introduced the partition functions
\bea
Z_\lambda(\Sigma_{\bm k}\,\hat\omega)  &=& \int \frac{\rd u}{\sqrt{2\pi/i}}~
\exp\left[-\frac{1}{2}\bigg(i\lambda_\varepsilon  + \Sigma_{\bm k}\,\hat\omega \bigg) u^2 \right] \nonumber\\
&=& \left(\frac{i}{i \lambda_\varepsilon + \Sigma_{\bm k}\,\hat\omega}\right)^{1/2}
\label{defZmu}\ ,\\
 Z_2(\omega,\omega',K) &=& \int \rd u \rd v ~
\exp\left[-\frac{1}{2}\bigg(\omega u^2 + \omega' v^2 - 2i K u v\bigg)\right] 
= \frac{2\pi}{\sqrt{\omega \omega' + K^2}}\ .
\label{defZ2}
\eea

The normalisation constant $\cN$ in (\ref{eq:avznd}) is given by
\bea
\cN&=&\int \prod_{i\nu} \frac{\rd \varphi_{i}^\nu}{2\pi} 
\exp\left\{\sum_\mu \frac{c^{\mu\mu}}{2N_\mu}\sum_{i\ne j\in \cN_\mu}\Big(\re^{\ri 
\varphi_i^\mu + \ri \varphi_j^\mu} -1 \Big) \right .\nonumber\\
&& \left .
 + \sum_{\mu<\nu} \frac{c^{\mu\nu}}{N_\nu}\sum_{i\in \cN_\mu}\sum_{j\in \cN_\nu}\Big( 
\re^{\ri \varphi_i^\nu + \ri \varphi_j^\mu}   -1 \Big) - \ri\sum_{\mu\nu} 
\sum_{i\in N_\mu}  \varphi_i^\nu k_i^\nu\right\}
\eea
Site decoupling is achieved by introducing
\be
\rho_0^{(\mu\nu)}= \frac{1}{N_\mu} \sum_{i\in\cN_\mu} \re^{\ri \varphi_i^\nu}
\ee
and a corresponding set of conjugate order parameters to enforce these definitions.
Note that for reasons to become clear below, we use a notation previously employed for 
normalisation factors of replicated densities. This duplication is intentional, as it 
reveals terms in the numerator and denominator of (\ref{ZnPint}) exhibiting the 
same exponential scaling in $N$ when evaluated at the saddle point, and hence cancel. 
We get
\bea
\cN&=&\int\prod_{\mu\nu}\frac{\rd\hat \rho_0^{(\mu\nu)}\rd\rho_0^{(\mu\nu)}}{2\pi/N_\mu}
\exp\left\{N\Bigg[\frac{1}{2}\sum_{\mu\nu}f_\mu c^{\mu\nu}\Big(\rho_0^{(\mu\nu)} 
\rho_0^{(\nu\mu)}-1 \Big)\right .\nonumber\\
&& \left . -\ri \sum_{\mu\nu} f_\mu \hat \rho_0^{(\mu\nu)} \rho_0^{(\mu\nu)} +
\sum_{\mu\nu}  f_\mu \Bigg( \Big\langle k^\nu \ln \ri \hat \rho_0^{(\mu\nu)} \Big
\rangle_\mu - \Big\langle \ln k^\nu !\Big\rangle_\mu \Bigg) \Bigg]\right\}\ ,
\label{NPint}
\eea
which is also evaluated by the saddle point method.

Before we derive the saddle point conditions, we note that the functions $G_1$, $G_2$ and 
$G_3$ in the numerator of (\ref{ZnPint}) contain both $\cO(1)$ and $\cO(n)$ contributions
in the $n\to 0$ limit, such that the integrand in the numerator contains both terms that 
scale exponentially in $N$ and in $Nn$. The denominator ($\cN$), however, scales 
exponentially in $N$. Since the ratio (\ref{ZnPint}) scales exponentially in $Nn$, 
the $\cO(1)$ contributions to $G_1$, $G_2$ and $G_3$ should cancel with those of $\cN$ at 
the saddle point, which is indeed the case.

Evaluating first the stationarity conditions for $G=G_1+G_2+G_3$ at $\cO(1)$ gives
\be
\ri \hat \rho_0^{(\mu\nu)} = c^{\mu\nu} \rho_0^{(\nu\mu)} \qquad \mbox{and}\qquad
\rho_0^{(\mu\nu)}=\frac{\langle k^\nu\rangle_\mu}{\ri \hat \rho_0^{(\mu\nu)}} = 
\frac{c^{\mu\nu}}{\ri \hat \rho_0^{(\mu\nu)}}
\ee
from which we obtain
\be
\ri \hat \rho_0^{(\mu\nu)} \rho_0^{(\mu\nu)} = c^{\mu\nu}  \qquad \mbox{and}\qquad
\rho_0^{(\mu\nu)}\rho_0^{(\nu\mu)} = 1\ .
\label{statnorm}
\ee
The stationarity conditions for the saddle point of $\cN$ are exactly the same. Since 
(\ref{statnorm}) exhibits the gauge-symmetry
\be
\rho_0^{(\mu\nu)} \to \rho_0^{(\mu\nu)} a^{(\mu\nu)} \ , \quad
\ri \hat \rho_0^{(\mu\nu)} \to \ri \hat \rho_0^{(\mu\nu)}/a^{(\mu\nu)}\ , \quad
\rho_0^{(\nu\mu)} \to \rho_0^{(\nu\mu)} / a^{(\mu\nu)} \ ,
\ee
the correct scaling in $Nn$ is obtained, provided that the same gauge is adopted in both 
the numerator and the denominator of (\ref{ZnPint}).
The saddle point contribution is determined from stationarity conditions with respect to 
variations of the $\pi^{(\mu\nu)}(\omega)$ and the $\hat\pi^{(\mu\nu)}(\hat \omega)$.

Using (\ref{statnorm}), the stationarity conditions for the $\pi^{(\mu\nu)}(\omega)$ read 
\be
\int \rd \hat\pi^{\mu\nu)}(\hat{\omega})
\ln \Bigg[\frac{Z(\hat{\omega}+\omega)} {Z(\hat{\omega})Z(\omega)}\Bigg]
= \int\rd\pi^{(\nu\mu)}(\omega')
\left\langle \ln \frac{Z_2(\omega,\omega',K)}{Z(\omega)Z(\omega')}\right\rangle_{\mu\nu}
+ \phi_{\mu\nu}
\label{statpi}
\ee
with $\phi_{\mu\nu}$ a Lagrange multiplier to enforce the normalisation of 
$\pi^{(\mu\nu)}$.\newline
The stationarity conditions for the $\hat\pi^{(\mu\nu)} (\hat{\omega})$ are
\be
c^{\mu\nu} \int \rd\pi^{(\mu\nu)}({\omega}) \ln \Bigg[\frac{Z(\hat{\omega}+\omega)} 
{Z(\hat{\omega})Z(\omega)}\Bigg] =\Bigg\langle k^\nu 
\int  \{\rd \hat\pi^{(\mu)}\}_{{\bm k}\setminus k^\nu}
\ln \frac{Z_\lambda \big(\hat{\omega} + \Sigma_{\bm k\setminus k^\nu}\,\hat\omega \big)}
{\{Z\}_{{\bm k}\setminus k^\nu}}\Bigg\rangle_\mu  + \hat \phi_{\mu\nu}
\label{stathatpi}
\ee
where $\{\rd \hat\pi^{(\mu)}\}_{{\bm k}\setminus k^\nu}$ denotes the product
$\{\rd \hat\pi^{(\mu)}\}_{\bm k}$ of integration measures from which $\rd\hat\pi^{(\mu\nu)}
(\hat \omega_{k^\nu})$  is excluded, i.e. the  product $\{\rd\hat\pi^{(\mu)}\}_{{\bm k}
\setminus k^\nu}\equiv\prod_{\tilde\nu(\ne\nu)}\prod_{\ell_{\tilde\nu}=1}^{k^{\tilde\nu}}
\rd\hat\pi^{(\mu{\tilde\nu})}(\hat\omega_{\ell_{\tilde\nu}})\times
\prod_{\ell_\nu=1}^{k^\nu-1}\rd\hat\pi^{(\mu\nu)}(\hat\omega_{\ell_\nu})$.
Analogous constructions apply to the product $\{Z\}_{{\bm k}\setminus k^\nu}$ and the sum 
$\Sigma_{\bm k\setminus k^\nu}\,\hat\omega$, and $\hat \phi_{\mu\nu}$ is the Lagrange 
multiplier to enforce the normalisation of the $\hat\pi^{(\mu\nu)}(\hat{\omega})$. 

Following \cite{Mon98,MezPar01}, the stationarity conditions for $\pi^{(\mu\nu)}(\omega)$ 
and $\hat\pi^{(\mu\nu)}(\hat\omega)$ are  rewritten in a form that suggests solving 
them via a population based algorithm.  In the present case we get \cite{Ku08,ErgKu09}
\bea
\hat\pi^{(\mu\nu)}(\hat{\omega})&=& \int\rd\pi^{(\nu\mu )}(\omega') 
\left\langle\delta(\hat{\omega} - \hat{\Omega}(\omega',K)\right\rangle_{\mu\nu}
\label{eq:pi}\\
\pi^{(\mu\nu)}({\omega}) &=& \Bigg\langle \frac{k^\nu}{c^{\mu\nu}} 
\int  \{\rd \hat\pi^{(\mu)}\}_{{\bm k}\setminus k^\nu} \delta\left(\omega - \Omega^{(\mu)}_{{\bm k}
\setminus k^\nu}\right) \Bigg\rangle_\mu
\label{eq:hatpi}
\eea
with
\be
\hat\Omega(\omega',K) = \frac{K^2}{\omega'}\ ,
\qquad \Omega^{(\mu)}_{{\bm k}\setminus k^\nu} = \ri\lambda_\varepsilon  + \Sigma_{\bm k\setminus 
k^\nu}\,\hat\omega \ .
\label{eq:Om-hatOM}
\ee

The spectral density is obtained from (\ref{defdos}); note that only the explicit $\lambda$
dependence in  $G_3$ in (\ref{G3n}) contributes. We obtain a formal result analogous to 
that obtained earlier for homogeneous systems, or for cross-connected modules of equal 
size with Poisson distributions of inter-module coordinations \cite{ErgKu09},
\be
\rho(\lambda)= \frac{1}{\pi}{\rm Re}\lim_{\varepsilon\searrow 0} \Big[\sum_\mu 
f_\mu q_d^{(\mu)}\Big]\ ,
\label{DOS}
\ee
where
\bea
q_d^{(\mu)} &=& \Bigg \langle \int \{\rd \hat\pi^{(\mu)}\}_{\bm k} ~ \Big\langle u^2\Big\rangle_{\{\hat\omega\}_{\bm k}}
\Bigg\rangle_{\mu}
\nonumber  \\
& = & \Bigg \langle \int \{\rd \hat\pi^{(\mu)}\}_{\bm k} ~\frac{1}{i \lambda_\varepsilon  + 
\Sigma_{\bm k}\,\hat\omega} 
\Bigg\rangle_{\mu}\ .
\label{stathatqd}
\eea
Here $\left\langle \dots \right\rangle_{\{\hat\omega\}_{\bm k}}$ is an average w.r.t. 
the Gaussian weight in terms of which $Z_\lambda(\Sigma_{\bm k}\,\hat\omega)$ is defined.

As explained in detail in \cite{Ku08}, an evaluation of (\ref{DOS}), (\ref{stathatqd}) via 
sampling from a population  will miss the pure-point contributions to the spectral density. 
In order to see these, a small non-zero regularizing $\varepsilon$, which amounts to replacing 
$\delta$-functions by Lorentzians of width $\varepsilon$ must be kept, resulting in a
density of states which is smoothed at the scale $\varepsilon$. A simultaneous evaluation 
of (\ref{stathatqd}) for non-zero $\varepsilon$ {\em and\/} in the $\varepsilon\searrow 
0$-limit then allows to disentangle pure-point and and continuous contributions to the 
total density of states (\ref{DOS}).

If we are interested in spectra of the generalised graph Laplacians $L$ defined by 
(\ref{ML}) instead of the weighted adjacency matrices $M$, we need to evaluate
\bea
\langle Z_N^n \rangle &=&\frac{1}{\cN}\  \int \prod_{i\nu} \frac{\rd \varphi_i^\nu}{2\pi}
\prod_{ia} \frac{\rd u_{ia}}{\sqrt{2\pi/\ri}} \exp\left\{-\frac{\ri}{2}\lambda_\varepsilon
\sum_{ia} u_{ia}^2 - \ri\sum_{\mu\nu} \sum_{i\in N_\mu}\varphi_i^\nu k_i^\nu \right .
\nonumber\\
&& \left .
+\sum_\mu\frac{c^{\mu\mu}}{2N_\mu}\sum_{i\ne j\in \cN_\mu}\left(~\Big\langle\exp\Big\{-\ri 
\frac{K}{2}\sum_{a}(u_{ia}-u_{ja})^2\Big\}\Big\rangle_{\mu\mu}\re^{\ri\varphi_i^\mu+\ri
\varphi_j^\mu}-1\right)\right .\nonumber\\
&& \left .
+\sum_{\mu<\nu} \frac{c^{\mu\nu}}{N_\nu}\sum_{i\in \cN_\mu}\sum_{j\in \cN_\nu} \left( 
~\Big\langle\exp\Big\{-\ri\frac{K}{2}\sum_{a} (u_{ia}-u_{ja})^2 \Big\}\Big\rangle_{\mu\nu} 
\re^{\ri \varphi_i^\nu + \ri \varphi_j^\mu}   -1 \right) 
\right\}
\label{eq:avzndL}
\eea
instead of (\ref{eq:avznd}), the only difference being the translationally invariant form 
of the interactions in the present case.\footnote{There is a typo, a missing minus-sign in
front of the translationally invariant interaction term in the last unnumbered Eq. on p 15
of \cite{Ku08}.}.
The structure of the theory developed above and the fixed point equations (\ref{eq:pi}), 
(\ref{eq:hatpi}) remain formally unaltered, apart from a modification of the definition of
$Z_2(\omega,\omega',K)$ of (\ref{defZ2}) due to the modified interaction term
\bea
Z_2(\omega,\omega',K) &=& \int \rd u \rd v ~
\exp\left[-\frac{1}{2}\bigg(\omega u^2 + \omega' v^2 + i K (u - v)^2 \bigg)\right] \nn\\
&=& Z(\omega'+\ri K)\, Z\bigg(\omega + \frac{K \omega'}{K-\ri\omega'}\bigg) \ ,
\label{defZ2n}
\eea
here expressed in terms of the normalisation constants $Z(\cdot)$ of (\ref{defZ}). 
As already noted in \cite{Ku08} this only requires a modified definition of
$\hat\Omega(\omega',K)$ in (\ref{eq:Om-hatOM}), viz
\be
\hat\Omega(\omega',K) = \frac{K \omega'}{K-\ri\omega'}\ ,
\label{LhatOm}
\ee
but leaves the the self-consistency equations otherwise unchanged.

\subsection{Cavity Equations for Finite Instances}
\label{sec:cav}
Rather than studying the ensemble in the thermodynamic limit, one can 
also look at large but finite single instances. The method of choice to study these 
is the cavity approach \cite{Rog+08}, for which the additional structure coming from 
modularity does not cause any additional complication at all, and the original set-up
\cite{Rog+08} applies without modification, apart from that related to generating 
large modular graphs with the prescribed statistics of inter- and intra-module 
connectivities. 

Equations (11) of \cite{Rog+08}, when written in terms of the notation and conventions 
used in the present paper translate into 
\be
\hat \omega_\ell^{(j)}= \frac{K_{j\ell}^2}{\omega_\ell^{(j)}}\ , \qquad 
\omega_j^{(i)}= \ri \lambda_\varepsilon + \sum_{\ell\in \partial j\setminus i} \hat 
\omega_\ell^{(j)}\ ,
\ee
in which $\partial j$ denotes the set of vertices connected to $j$, and $\partial j
\setminus i$ the set of neighbours of $j$, excluding $i$.

These equations can be solved iteratively \cite{Rog+08} even for very large system sizes, 
showing fast convergence except at mobility edges, where we observe critical slowing-down.
The density of states for a single instance of a matrix $M$ is obtained from the 
self-consistent solution via  
\be
\rho_M(\lambda) = \lim_{\varepsilon\searrow 0} \frac{1}{N\pi} \sum_j {\rm Re }
\left[\frac{1}{\omega_j}\right]
\ee
with
\be
\omega_j= \ri \lambda_\varepsilon + \sum_{\ell\in \partial j} \hat \omega_\ell^{(j)}\ .
\ee
The modifications required to treat generalised graph Laplacians are once more 
straightforward.

\section{Small-World Networks}
\label{sec:smw}
Small-world networks can be constructed from any graph, by introducing a second,
random connectivity component which introduces short-cuts in the original graph, as 
long as the second component is sufficiently weakly correlated with the first. 

The standard example is a closed ring, with additional links between randomly chosen 
pairs along the ring. Alternatively one could start with a regular random graph of 
fixed coordination $k_i=2$ (this gives an ensemble of loops with typical lengths diverging
in the thermodynamic limit $N\to\infty$), then introducing a second 
sparse connectivity component linking randomly chosen vertices of the original graph.

Clearly the original graph need not be a ring; it could be a $d$ dimensional lattice, 
a Bethe lattice, or a (regular) random graph of (average) connectivity different from 2, 
and one could introduce {\em several\/} additional random connectivity components to 
create short-cuts. 

In what follows, we look at (finitely coordinated) random graphs with several 
connectivity components between the vertices of the graphs. The set-up is rather close
to that of multi-modular systems as described above, except that there is only a single
module, having $m$ connectivity components linking the vertices of this single module.

The formal structure of the theory is therefore very similar to that described earlier and
we just quote the final fixed point equations, and the result for the spectral density,
without derivations.

We need to solve the following set of fixed point equations
\bea
\hat\pi^{(\nu)}(\hat{\omega})&=& \int\rd\pi^{(\nu)}(\omega') 
\left\langle\delta(\hat{\omega} - \hat{\Omega}(\omega',K)\right\rangle_{\nu}
\label{smw:pi}\\
\pi^{(\nu)}({\omega}) &=& \Bigg\langle \frac{k^\nu}{c^{\nu}} 
\int  \{\rd \hat\pi\}_{{\bm k}\setminus k^\nu} \delta\left(\omega - \Omega_{{\bm k}
\setminus k^\nu}\right) \Bigg\rangle
\label{smw:hatpi}
\eea
with
\be
\hat\Omega(\omega',K) = \frac{K^2}{\omega'}\ ,
\qquad \Omega_{{\bm k}\setminus k^\nu} = i\lambda_\varepsilon  + \Sigma_{\bm k\setminus k^\nu}
\,\hat\omega \ .
\label{smw:Om-hatOM}
\ee
where now $\langle \dots\rangle_\nu$ in (\ref{smw:pi}) denotes an average over the weight 
distribution of the $\nu$-th coupling component $\{K_{ij}^{(\nu)}\}$, and the average 
$\langle \dots\rangle$ in (\ref{smw:hatpi}) is over the distribution of $m$-dimensional 
coordinations $\bm k_i=(k_i^\nu)$, with $\langle k_i^\nu \rangle = c^\nu$. The (average) 
spectral density is then given by
\be
\rho(\lambda) = \frac{1}{\pi}{\rm Re}~\lim_{\varepsilon\searrow 0}~\Bigg \langle 
\int \{\rd \hat\pi\}_{\bm k} ~\frac{1}{i \lambda_\varepsilon  + \Sigma_{\bm k}\,\hat\omega} 
\Bigg\rangle\ .
\ee

Were one to look at the graph Laplacian for this type of small-world network, rather
than at weighted) adjacency matrices one would once more only have to substitute 
(\ref{LhatOm}) for $\hat \Omega$ in (\ref{smw:Om-hatOM}), as discussed in for the 
multi-modular case above.

\section{Results}
\label{sec:results}
\subsection{Modular Systems}
For the multi-modular systems, there are clearly far too many possible parameters and 
parameter combinations to even begin to attempt giving an overview of the phenomena 
one might see in such systems. Hence, we restrict ourselves to just one illustrative
example chosen to highlight how the total density of states in different parts of the 
spectrum may be dominated by contributions of local densities of states of specific 
sub-modules.

\begin{figure}[t]
$$\epsfig{file=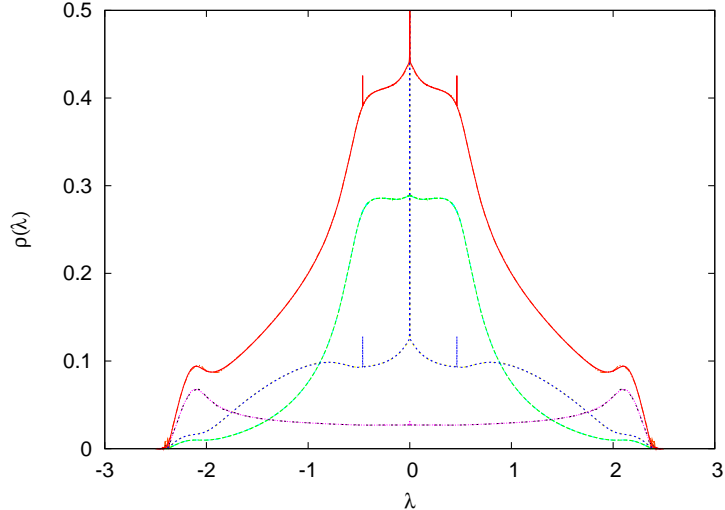,width=100mm}$$
\caption{(Colour online) Spectrum of the three-module system described in the main text 
(red full line), along with its unfolding according to contributions of local densities of 
states contributed by the three sub-modules as suggested by Eq (39) (green long dashed: 
module 1, blue short dashed: module 2, and magenta dot-dashed: module 3). Ensemble results
are displayed together with the corresponding results for a single instance of size 
$N=64,000$. The former are plotted on top of the latter, so colour coded lines for the 
finite instance results (light blue: module 1, yellow: module 2, black dashed: module 3, 
and light-red dashed for the total DOS) remain basically invisible due to the nearly perfect 
match.}
\end{figure}

We present a system consisting of three modules, with fractions $f_1=1/2$, $f_2=1/3$, and
$f_3=1/6$ of the system respectively. Modules 1 and 3 have fixed intra-modular
connectivities with coordinations 3 and 2 respectively, while module 2 has Poisson 
connectivity with average 2. Inter-module connectivities are all Poissonian with averages 
$c^{12}=1$ and $c^{13}=c^{23}=1/2$ ($c^{21},~c^{31}$ and $c^{32}$ follow from the 
consistency requirements). Non-zero couplings are chosen bi-modal $\pm 1/\sqrt c_t$ 
with $c_t = \sum_{\mu\nu} f_\mu c^{\mu\nu}$ apart from intra-module couplings in modules 
1 and 3, which have values of ${\pm 1/2\sqrt c_t}$ and ${\pm 2/\sqrt c_t}$ respectively.

Figure 1 shows the results for this system. We observe that the central cusp at $\lambda=0$
and the $\delta$-function contributions to the total density of states at $\lambda=0$ and
at $\lambda \simeq \pm 0,463$ essentially originate from module 2 (with Poisson connectivity 
of average coordination 2); a regularizing $\varepsilon = 10^{-4}$ has been used to exhibit
the $\delta$-function contributions. The humps at the edges of the spectrum mainly come from 
module 3 with the fixed coordination 2, whereas the shape of the shoulders at intermediate 
$\lambda$ values are mostly determined by the largest module 1 with fixed coordination 3.
Note that there are small tails of localised states for $|\lambda| \gtrsim 2.415$.

We found results computed for a single instance of this modular structure containing $N=60,000$
vertices to be virtually indistinguishable from the ensemble results, except for finite sample 
fluctuations in the extreme tails where the expected DOS becomes too small to expect more 
than a few eigenvalues for the $N=60,000$ system.

The $\delta$-peaks at $\lambda \simeq \pm 0,463$ originate from isolated dimers as part of 
module 2 which remain isolated upon cross-linking the different modules. For the three 
module system in question, the weight $a_2$ of each of the $\delta$-peaks can be shown to be 
$\frac{1}{6}\re^{-8} \simeq 5.6\,10^{-5}$ in the thermodynamic limit, compatible with a 
rough estimate of $a_2 = (5 \pm 1) \, 10^{-5}$ from our numerical ensemble results. We note 
in passing that pure-point contributions to the spectral density would be generated by many 
other finite isolated clusters; the ones with the next highest weight would be generated 
by isolated open trimers, but these are more than an order of magnitude less likely to 
occur, so that we have not picked them up at the precision with which we have performed 
the $\lambda$ scan in Fig. 1.

\subsection{Small-World Networks}

The small-world networks we consider here have a small fraction of long-range connections 
added to a regular random graph of fixed coordination 2. The system {\em without\/} 
long-range interactions is effectively an infinite ring; it can be diagonalised 
analytically; for couplings of unit strength it has a band of extended states for 
$|\lambda|\le 2$, and the density of states exhibits the typical integrable van Hove 
singularity $\rho(\lambda)\sim \big | \, |\lambda|-2 \big|^{-1/2}$ of a one-dimensional 
regular system.

When a small amount of weak long-range interactions is introduced into the system, this 
central band will initially slightly broaden, and the van Hove singularity gets rounded 
(the integrable divergence disappears). As the strength of the long-range connections is
increased, the central band is widened further. At the same time the density of states 
acquires some structure,  which becomes more intricate, as the strength of the long-range
interactions is increased, including side-peaks which themselves acquire sub-structure, 
and typically a depression of the DOS near the location of the original band edge. This 
depression deepens with increasing strength of the long-range interactions, and eventually
becomes a proper band-gap, which we find to be populated only by {\em localised\/} states.
Further increase in interaction strength introduces ever more structure, including  
depressions in the DOS inside side-bands which in turn develop into proper band gaps.

Figure 2 shows a system for which the average additional long-range coordination is 
$c=0.5$,  so that long-range interactions are associated with fewer than half of the nodes
on the ring. The figure displays the central region of the spectrum for a range of 
interaction strengths of the long-range couplings, which exhibit increasing amounts of 
structure with increasing interaction strengths.

\begin{figure}[ht]
$$\epsfig{file=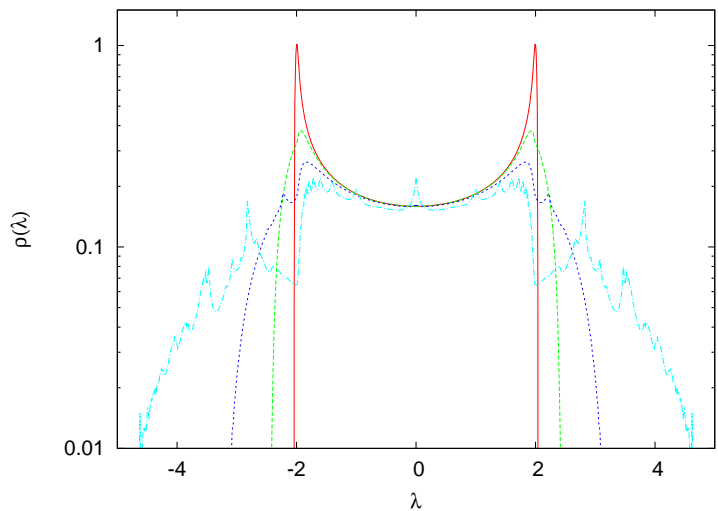,width=100mm}$$
\caption{(Colour online) Central part of the spectrum the small-world system described in
the text. The four curves correspond to long-range interactions of strengths $J=0.1$ (red 
full line), $J=0.5$ (green long dashed), $J=1.0$ (blue short dashed), and $J=2.0$ (magenta 
dot-dashed).}
\end{figure}

Figure 3 shows the entire spectrum of this system at  $J=5$, and separately exhibits the 
contribution of the continuous spectrum. Now 4 distinct separate side bands of continuous 
states can clearly be distinguished on each side of the central band, with proper band 
gaps (filled with localised states) between them. 

The spectrum shown in Figure 3 displays structure at many levels. To mention just two of 
the more prominent ones: the original edge of the central band develops a sequence of peaks
which extends into the localised region. Side-bands too acquire multi-peak structures, with
individual peaks exhibiting further sub-structure. 

Subject to limitations of computational power, our algorithm is able to exhibit these
structures to any desired level of accuracy, though in some regions --- predominantly at 
band edges --- our data for the continuous DOS remain somewhat noisy; we suspect that in 
such regions there is a set of localised states that becomes dense in the thermodynamic 
limit, which is responsible for this phenomenon. Also, we have a localisation transition at
every band-edge which may well induce critical slowing down in the population dynamics
algorithm by which we obtain spectral densities. Quite possibly because of this, finite
population-size effects in the population dynamics are much stronger in the present 
small-world system than in the simpler systems without side-bands studied before 
\cite{Ku08, Rog+08, ErgKu09}.

We have attempted to verify the localisation transitions using numerical diagonalisation
and computations of inverse participation ratios \cite{EvangEcon92, Evang92} in finite 
instances of increasing size, but the convergence to asymptotic trends is extremely slow.
Although we have gone to system sizes as large as $N=3200$ for this system, the numerical
results, while compatible with those derived from our population dynamics algorithm, are 
still not forceful enough to strongly support them. These aspects clearly deserve further
study. In this respect a recent result of Metz et al. \cite{Metz+10}, who managed to 
compute IPRs within a population dynamics approach, could well provide the method of 
choice to clarify the situation, though we have not yet implemented their algorithm. 

\begin{figure}[ht]
$$\epsfig{file=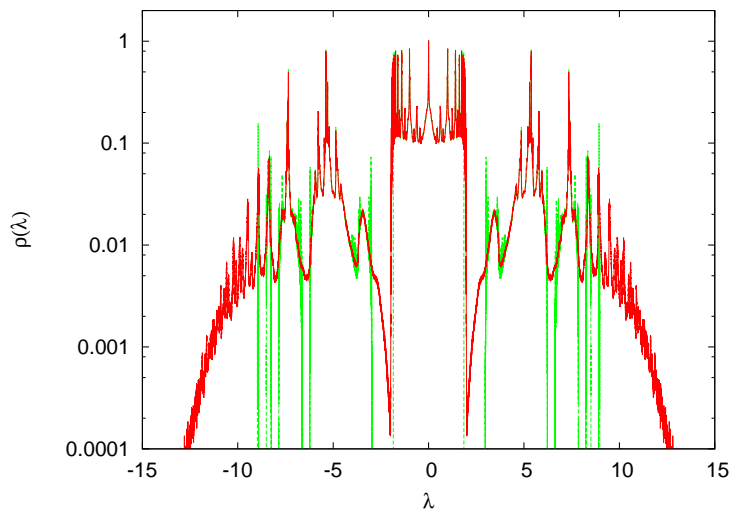,width=100mm}$$
\caption{(Colour online) Spectrum of the small-world system with $c=0.5$ at $J=5$, showing
both the DOS of the extended states (green long dashed line), and the total DOS including 
contributions from localised states (red full line). A regularizing $\varepsilon=10^{-3}$
has been used to exhibit the latter.}
\end{figure}

\subsection{Graph Laplacians}

From a dynamical point of view, Graph Laplacians (\ref{ML}) are in many ways more
interesting than the corresponding connectivity matrices (\ref{M1}), as they could be used
to analyse e.g. diffusive transport on graphs, to give vibrational modes of structures 
described by these graphs, or to define the kinetic energy component of random 
Schr\"odinger operators. We have accordingly also looked at spectra of the graph Laplacian 
corresponding to the small-world type structures discussed in the previous section.

For the regular random graph of fixed coordination 2, the spectrum of the graph Laplacian
is just a shifted version of the spectrum of the connectivity matrix. As for the latter, 
by adding a small amount of weak long-range interactions this translated band initially 
broadens slightly, and the van Hove singularities disappear. More or stronger long-range 
interactions do, however, not appear to create much structure in the initial band. The 
tails at the lower band edge do acquire structure, and eventually develop proper band-gaps,
populated only by localised states, much as for the connectivity matrix. 

Figure 4 shows the spectrum of the graph Laplacian for a small-world system with the same
parameters as in Fig. 3. We recognise a main band of extended states for $-5.59\lesssim 
\lambda\lesssim -0.037$ and four side bands, two of which are very narrow; they are centred 
around $\lambda\simeq -7.2$ and $\lambda\simeq -7.3$, and are barely distinguishable as separate 
bands on the scale of the figure. Although the data for $\lambda < -20$ appears to look
noisy, the fine structure in this region of the spectrum is actually accurate; as shown in
the inset, we found them to be very well reproduced by high precision exact diagonalisation
of an ensemble of $10^4$ matrices of size $1600\times1600$, using a fine binning (5000 bins
across the entire spectrum, thus $\Delta \lambda \simeq 8 \,10^{-3}$), to achieve sufficient 
resolution of details. 

\begin{figure}[h!]
\setlength{\unitlength}{1mm}
\begin{picture}(200,130)(0,0)
\put(-10,5){\epsfig{file=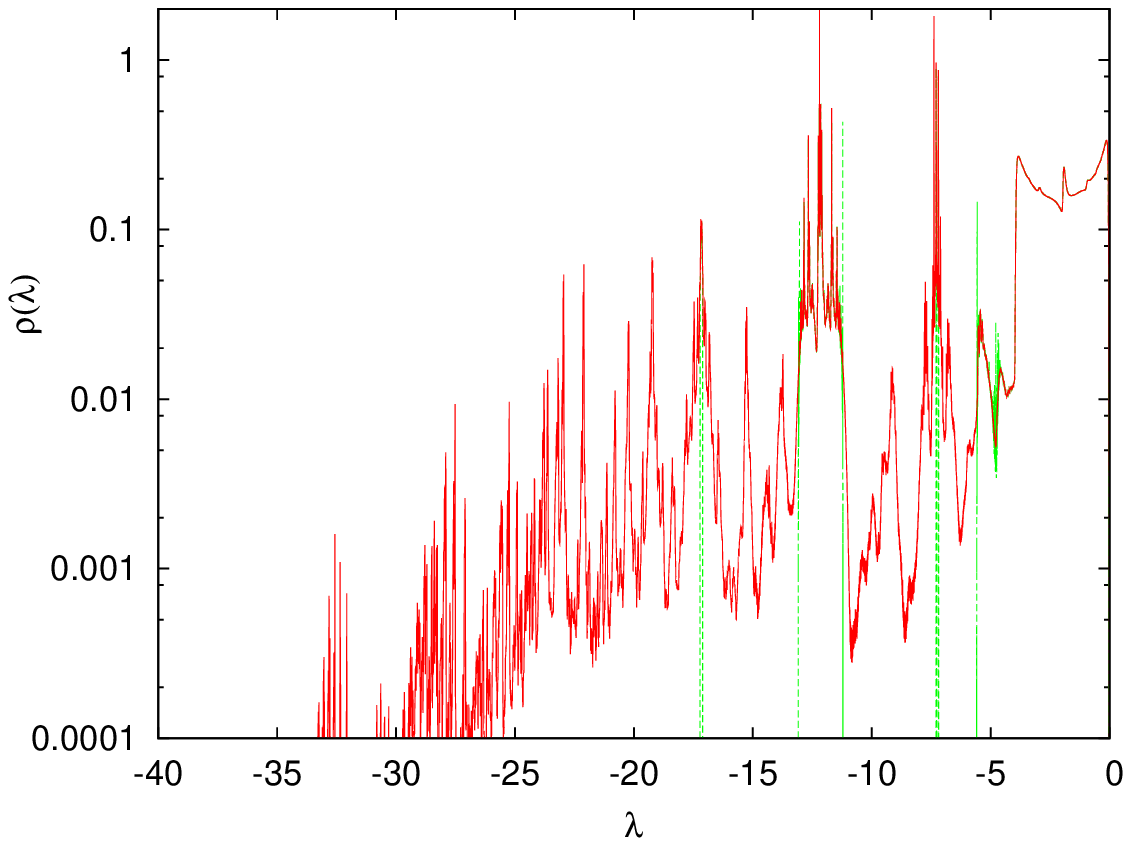,width=170mm}}
\put(25,80){\epsfig{file=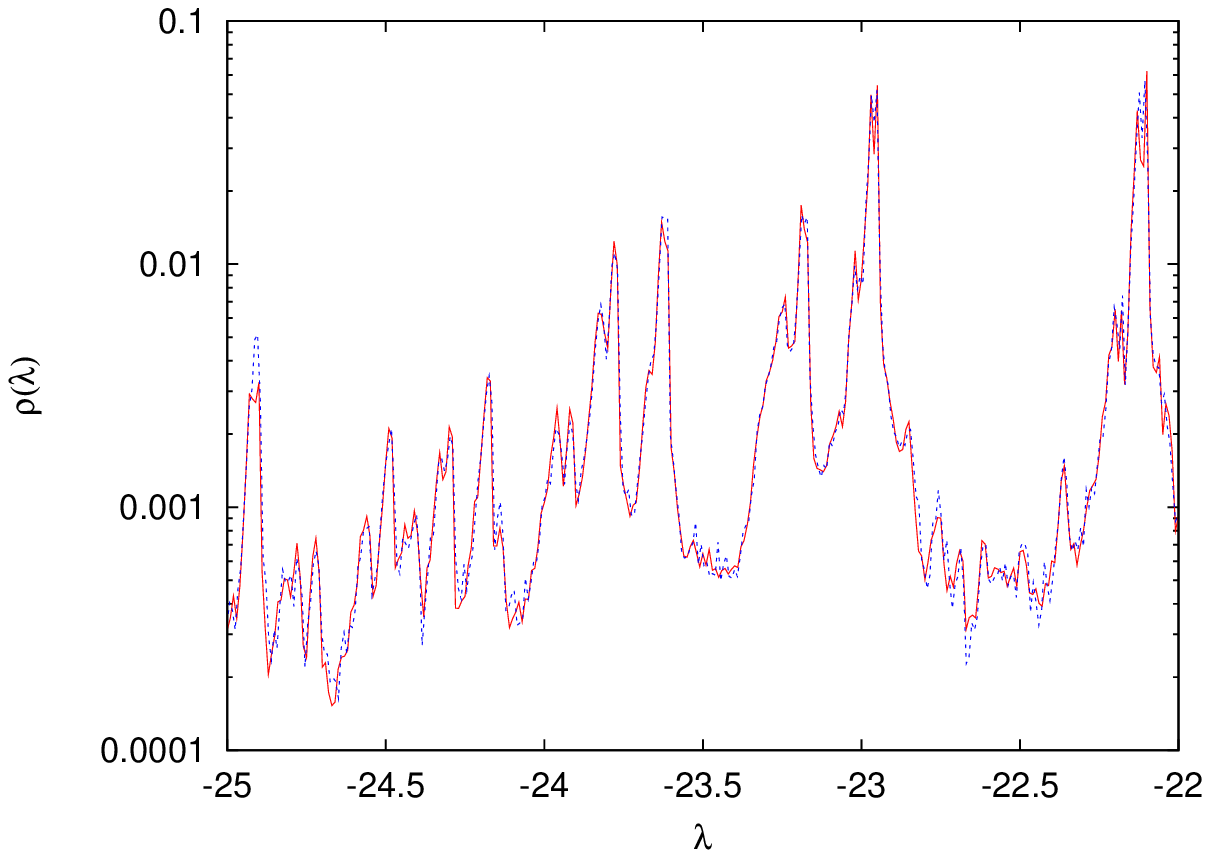,width=52mm}}
\end{picture}
\caption{(Colour online) Spectrum the graph Laplacian for a small-world system with 
$c=0.5$ at $J=5$, showing both the DOS of the extended states (green long dashed 
line), and the total DOS including contributions from localised states  (red full line),
regularized at $\varepsilon=10^{-3}$. The inset compares the population dynamics results 
for the total DOS (red full line) with results of direct numerical diagonalisation (blue 
short dashed line) in the range $-25\le\lambda\le -22$.}
\end{figure}

The appearance of several bands of extended states, separated by gaps which are populated 
only by localised states implies that transport processes such as diffusion will exhibit 
several distinct time-scales for such systems. Given the way in which the system is 
constructed, the appearance of two time-scales would not be surprising, as diffusion takes
place both along the ring and via short cuts. The fact that there are several such time 
scales would not seem obvious, though.

Another feature which becomes apparent only by zooming into the region of very small 
$|\lambda|$ is the appearance of a mobility edge at $\lambda_c \simeq -0.037$ and a region
of localised states for $\lambda>\lambda_c$. The behaviour of the spectral density in the 
localised region $\lambda_c\le\lambda\le0$ shows singular Lifshitz type behaviour 
\cite{Khor+06}. For systems with a range of different parameters, both for the average 
number $c$ of long-range connections per site, and for their strength $J$, we find it to 
be compatible with the functional form
\be
\rho(\lambda) \simeq a \exp(-b/|\lambda|^{2/3})\ ,
\label{fit}
\ee 
with $a$ and $b$ depending on $c$ and $J$. For the $c=0.5$, $J=5$ system shown in Fig. 4 
we have $a=4.0 \pm 0.1$ and $b=0.49 \pm 0.003$. Three parameter fits which attempt to 
determine the $|\lambda|$ power in the exponential of (\ref{fit}) do give powers slightly 
different from $2/3$ at comparable values of reduced $\chi^2$, but the uncertainties of 
individual parameters are much larger. It may be worth mentioning that we have observed 
similar Lifshitz tails also for Laplacians of simple Poisson random graphs, both below 
and above the percolation transition, and for Laplacians corresponding to modular random 
graphs such as the one studied in Sect 4.1.

\begin{figure}[h!]
$$\epsfig{file=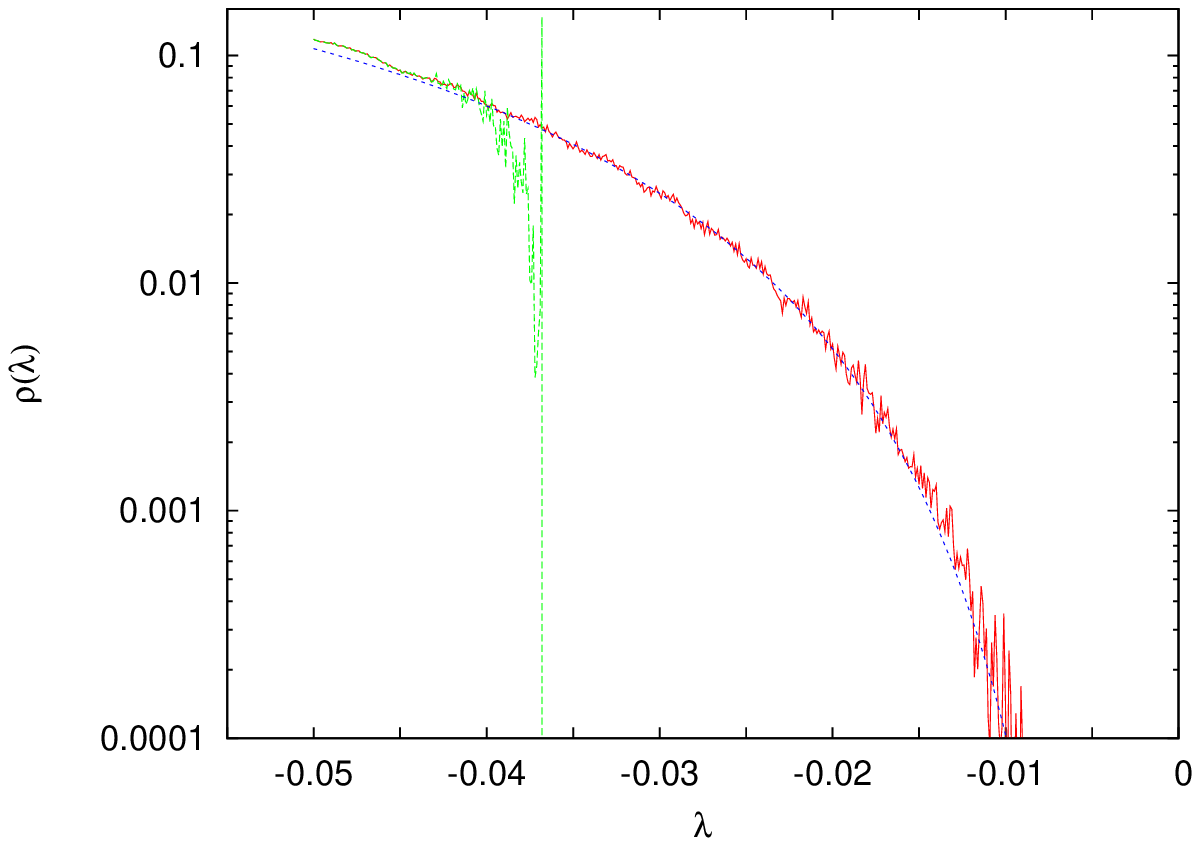,width=100mm}$$
\caption{(Colour online) Lifshitz tail for small $|\lambda|$. Shown are the band edge of 
the extended states (green long dashed line), the total DOS including contributions from 
localised states (red full line), and (\ref{fit}) (blue short dashed), based on a fit of 
the data in the interval $-0.035 \le \lambda \le -0.0075$.}
\end{figure}

\section{Conclusions}
\label{sec:concl}
We have computed spectra of matrices  describing random graphs with modular or small-world 
structure, looking both at connectivity matrices and at (weighted) graph Laplacians. 
Spectra are evaluated for random matrix ensembles in the thermodynamic limit using replica, 
and for large single instances using the cavity method. We find excellent agreement between 
the two sets of results if the single instances are sufficiently large; graphs containing 
$N ={\cal O}(10^4 -10^5)$ vertices are typically required to achieve agreement with relative
errors below $10^{-3}$. The ensemble and single instance results are in turn in excellent
agreement with results of direct numerical diagonalisations, though averages over many 
samples are required for the latter due to the comparatively moderate sample sizes that 
can be handled in the direct diagonalisation approach.

For a multi-modular system we have seen by way of example, how the total density of states 
in different parts of the spectrum may be dominated by contributions of local densities of 
states of specific sub-modules. The ability to identify such contributions may well
become a useful diagnostic tool in situations where one needs to study the topology of 
modular systems for which plausible null-models of their compositions are available.

For small-world systems, we have seen how the introduction of short-cuts in a regular
graph adds structure to the spectrum of the original regular random graph from which the
small-world system is derived. Depending on parameters, this may include the possibility
of having one or several satellite bands of extended states separated from the original 
band by gaps that are populated only by localised states. Whenever this happens for 
(weighted) graph Laplacians this implies the introduction of different time-scales for 
diffusive transport described by these Laplacians.

For graph Laplacians we typically observe a region of localised states at {\em small\/}
$|\lambda|$ where the density of states exhibits singular Lifshitz type behaviour. We
note that the existence of a small $|\lambda|$ mobility edge implies that these systems
will exhibit a {\em finite\/} maximum relaxation time for global diffusive modes
though there is no corresponding upper limit for the relaxation times for local modes.

We iterate that our methods are completely general concerning the modular structure of
the matrices. Concerning connectivity distributions, the only requirements are that
they are maximally random subject only to the constraints coming from prescribed degree
distributions. Modular graphs with additional topological constraints beyond degree 
distributions could be handled by suitably adapting the techniques of \cite{Rog+10}.

\bibliography{../../MyBib}
\end{document}